\documentclass[preprint,aps,amsmath,tightenlines,nofootinbib,showpacs]{revtex4}

\usepackage{epsfig}
\usepackage{amssymb}

\def\OMIT#1{{}}

\newcommand{\beq}{\begin{equation}}
\newcommand{\eeq}{\end{equation}}
\newcommand{\beqa}{\begin{eqnarray}}
\newcommand{\eeqa}{\end{eqnarray}}

\newcommand{\Choose}[2]{{\begin{pmatrix} {#1} \\ {#2} \end{pmatrix}}}

\begin{document}

\preprint{\vbox{
\hbox{UNH-10-01}}}

\title{Ground state energy of the interacting Bose gas\\ in two dimensions: an explicit construction}
\vskip 0.2cm
\author{\bf Silas R.~Beane}
\affiliation{Department of Physics, University of New Hampshire,
Durham, NH 03824-3568.}

\vphantom{}
\vskip 1.4cm
\begin{abstract} 

\noindent The isotropic scattering phase shift is calculated for
non-relativistic bosons interacting at low energies via an arbitrary
finite-range potential in $d$ spacetime dimensions.  Scattering on a
$(d-1)$-dimensional torus is then considered, and the eigenvalue
equation relating the energy levels on the torus to the scattering
phase shift is derived. With this technology in hand, and focusing on
the case of two spatial dimensions, a perturbative expansion is
developed for the ground-state energy of ${\bf N}$ identical bosons
which interact via an arbitrary finite-range potential in a finite area.
The leading non-universal effects due to range corrections and
three-body forces are included.  It is then shown that the
thermodynamic limit of the ground-state energy in a finite area can be
taken in closed form to obtain the energy-per-particle in the
low-density expansion, by explicitly summing the parts of the
finite-area energy that diverge with powers of ${\bf N}$.  The leading
and subleading finite-size corrections to the thermodynamic limit
equation-of-state are also computed.  Closed-form results --some
well-known, others perhaps not-- for two-dimensional lattice sums are
included in an appendix.

\end{abstract}


\maketitle

\tableofcontents

\vfill\eject

\section{Introduction}

\noindent The study of quantum mechanical scattering in a confined
geometry is topical in several distinct ways. Recently developed
experimental techniques involving trapped ultracold atoms are able to
alter spatial dimensionality~\cite{Hadz,clade,posa,Bloch:2008zz}, thus
motivating an understanding of the quantum mechanical interactions
among atoms as the number of spatial dimensions are continuously
varied. Bose gases in two spatial dimensions are of particular
interest as they are expected to have a complex phase structure which
is quite distinct from their counterparts in three spatial
dimensions~\cite{Mermin:1966fe,Hohenberg:1967zz,Kosterlitz:1973xp,fishe}. On the other
hand, from the perspective of numerical simulation, scattering in a
confined geometry is often a practical necessity. For instance, in
lattice studies of quantum field theories, calculations are done in a
four-dimensional Euclidean space time volume. For reasons of cost, the
finite spatial and temporal extent of these volumes is currently not
enormous as compared to the physical length scales that are
characteristic of the particles and interactions that are
simulated. Moreover, there are no-go theorems~\cite{Maiani:ca} for
Euclidean quantum field theory that require a finite volume
in order to extract information about hadronic interactions 
away from kinematic thresholds. The technology required to relate
hadronic interactions to the finite-volume singularities that are
measured on the lattice has been developed in 
Refs.~\cite{Luscher:1986pf,Luscher:1990ux,Beane:2003da,Beane:2007qr,Tan:2007bg,Detmold:2008gh}
and state-of-the-art Lattice QCD calculations have now measured the
energy levels of up to twelve interacting pions and allowed a determination
of the three-pion interaction~\cite{Beane:2007es}.
Similarly, quantum Monte Carlo studies of many-body systems in nuclear
and condensed matter physics are carried out in a finite volume or a
finite area, and thus a detailed understanding of how the energy
levels in the confined geometry map onto continuum physics is
essential to controlled predictive power.

The purpose of the present study is several-fold. First, we aim to
present a general study of the ground-state energy of a system of
${\bf N}$ bosons interacting via the most general finite-range
potential, confined to a finite area. This energy admits a
perturbative expansion in the two-body coupling strength for the case
of a weak repulsive interaction.  As a necessary prelude to
considering a confined geometry, we first review the subject of
isotropic scattering of identical bosons in $d$ spacetime dimensions
using effective field theory (EFT). It is assumed that the reader is
aware of the advantages of EFT technology. We then present a general
study of the relation between eigenenergies on a torus and
continuum-limit isotropic scattering parameters, for any spacetime
dimension. While the eigenvalue equation that we obtain is derived in
the non-relativistic EFT, it is expected to be generally valid in an
arbitrary quantum field theory up to corrections that are
exponentially suppressed in the size of the geometric boundary.  A
general study along these lines in quantum field theory is quite
involved and has been carried out only in four spacetime
dimensions~\cite{Luscher:1990ux}. The results of that study
demonstrated that boundary effects due to polarization are suppressed
exponentially with spatial size and therefore the leading power law
behavior can be found directly in the non-relativistic theory. Hence
the leading effects are captured by the non-relativistic EFT, with
relativistic effects appearing
perturbatively~\cite{Beane:2007qr,Detmold:2008gh}.  In the case of two
spatial dimensions, the exact two-body eigenvalue equation has been
considered previously in the context of lattice QED simulations in
three spacetime dimensions~\cite{Fiebig:1994qi}. However, there is
little discussion in the literature about the consequences of scale
invariance in the confined geometry, and about the ground-state energy of
the many-body system in a finite area. Moreover, to our
knowledge, the closed-form results that exist for the relevant lattice
sums in even spatial dimensions, which render this case a
particularly interesting theoretical playground, have not been noted
previously.

Following our derivation of the ground-state energy of a system of
${\bf N}$ bosons confined to a finite area, we demonstrate that
the thermodynamic limit of this system may be taken explicitly, by
summing the parts of the expansion that diverge in the large ${\bf N}$
limit. In the thermodynamic limit, the energy-per-particle admits a
double perturbative expansion in the two-body coupling and the
density. As a byproduct of taking the thermodynamic limit, we are able
to compute finite-size corrections to the thermodynamic-limit formula.  In
addition, we trivially include the leading non-universal corrections due
to three-body forces. Study of the weakly interacting Bose gas at zero
temperature has a long history, beginning with the mean-field result
of Schick~\cite{schic}, with subleading corrections computed in
Refs~\cite{popo,fishe,chern,Andersen}.  There are some
claims in the literature regarding discrepancies among the various
studies. We will comment on these claims below.

This paper is organized as follows. In section II we review low-energy
non-relativistic scattering of bosons in the continuum.  Using EFT we
calculate the isotropic phase shift in an arbitrary number of
spacetime dimensions. Section III considers low-energy
non-relativistic scattering of bosons in a confined geometry, in
particular on a $d-1$-dimensional torus.  We obtain the exact
eigenvalue equation which relates the energy levels on the torus to
the two-body continuum-limit scattering parameters.  In section IV we
consider the ground-state energy of a system of Bosons confined to a
finite area. We first develop perturbation theory on the
$d-1$-dimensional torus and recover the perturbative expansions of the
two-body results found previously. We then focus on ${\bf N}$ bosons
interacting via weak repulsive interactions in a finite area and give
a general expression for the ground-state energy.  In section V we
demonstrate how to take the thermodynamic limit in order to recover
the well-known low-density expansion, and we compare our results with
other calculations. We also compute the leading and subleading
finite-size corrections to the thermodynamic-limit energy
density. Finally, in section VI we conclude. In two Appendices, we
make use of some well-known exact results for even-dimensional lattice
sums to derive some closed-form expressions that are useful for the
case of two spatial dimensions, and we evaluate several sums involving
the Catalan numbers, which are relevant for deriving the
thermodynamic-limit equation-of-state.

\section{Scattering in the continuum}

\subsection{Generalities}

\noindent Here we will review some basic EFT technology which will
allow us to obtain a general expression for the isotropic scattering
phase shift in any number of dimensions.  If one is interested in
low-energy scattering, an arbitrary interaction potential of finite
range may be replaced by an infinite tower of contact operators, with
coefficients to be determined either by matching to the full theory or 
by experiment.  At low energies only a few of the contact operators will be
important. The EFT of bosons~\footnote{For a review, see Ref.~\cite{Braaten:2000eh}.}, destroyed by the field operator
$\psi$, which interact through contact interactions, has the following
Lagrangian:
%
\begin{figure}[t]
\includegraphics*[width=0.85\textwidth]{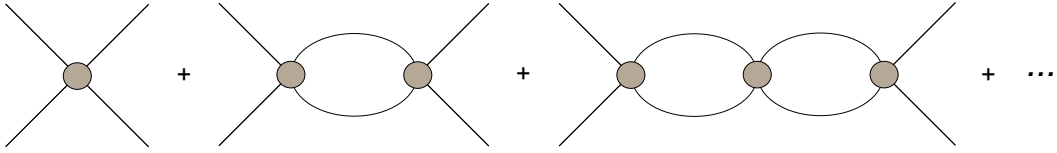}
\caption{\it Feynman diagrams that give the exact two-body scattering amplitude. The oval blob represents
the all-orders interaction derived from the Lagrangian.}
\label{fig:feynman}
\end{figure}
\beq
{\cal L}=
{{\psi}^\dagger} \left( i\partial_t + \frac{\nabla^2}{2M}\right) \psi
-\frac{C_0}{4} ({{\psi}^\dagger} \psi)^2
- \frac{C_2}{8} \nabla({{\psi}^\dagger } \psi)\nabla({{\psi}^\dagger } \psi) 
- \frac{D_0}{36} ({{\psi}^\dagger} \psi)^3\ +\ \ldots
\label{eq:1}
\eeq This Lagrangian, constrained by Galilean invariance, parity and
time-reversal invariance, describes Bosons interacting at low-energies
via an arbitrary finite-range potential. In principle, it is valid in
any number of spacetime dimensions, $d$. The mass dimensions of the
boson field and of the operator coefficients change with spacetime
dimensions: i.e.  $[\psi]=(d-1)/2$, $[C_{2n}]=2-d-2n$ and
$[D_{0}]=3-2d$.  While our focus in this paper is on $d=3$, in our
general discussion of two- and three-body interactions, we will keep
$d$ arbitrary as this will allow the reader to check our results
against the well-known cases with $d=2$ and $d=4$. Throughout we use
units with $\hbar =1$, however we will keep the boson mass, $M$,
explicit.

Consider $2\rightarrow 2$ scattering, with incoming momenta
labeled ${\bf p}_1,{\bf p}_2$ and outgoing momenta labeled ${\bf
  p}'_1,{\bf p}'_2$. In the center-of-mass frame, ${\bf p}={\bf
  p}_1=-{\bf p}_2$ , and the sum of Feynman diagrams, shown in
fig.~\ref{fig:feynman}, computed in the EFT gives the
two-body scattering amplitude~\cite{Braaten:2000eh,Kaplan:1998we,vanKolck:1998bw}
\begin{eqnarray}
{\cal A}_2(p) & = & -{ \sum C_{2n} \ p^{2n}  \over
1 - I_0(p) \sum C_{2n} \ p^{2n}} \ ,
\label{eq:2}
\end{eqnarray}
where
\begin{eqnarray}
I_0(p) \ = \ \frac{M}{2}\left({\mu\over 2}\right)^{\epsilon} \int {d^{D-1}{\bf q}\over
  (2\pi)^{D-1}}
{1\over p^2-{{\bf q}^2} + i \delta}
\ ,
\label{eq:2b}
\end{eqnarray}
and it is understood that the ultraviolet divergences in the EFT
are regulated using dimensional regularization (DR).
In eq.~(\ref{eq:2b}), $\mu$ and $D$ are the DR scale and dimensionality, respectively,
and $\epsilon\equiv d-D$. A useful integral is:
\begin{eqnarray}
\openup3\jot
I_n(p)&=& \frac{M}{2}\left({\mu\over 2}\right)^{\epsilon} \int {{{\rm d}}^{D-1}{\bf  q}\over (2\pi)^{D-1}}\, 
{\bf q}^{2n} \left({1\over p^2  -{\bf q}^2 + i\delta}\right) \ ;
\nonumber\\
&=& -\frac{M}{2} p^{2n} (-p^2-i\delta)^{(D-3)/2 } \Gamma\left({3-D\over 2}\right)
{(\mu/2)^{\epsilon}\over  (4\pi)^{(D-1)/2}}\ .
\label{eq:2a}
\end{eqnarray}
In what follows we will define the EFT coefficients in DR with
$\overline{MS}$. This choice is by no means generally
appropriate~\cite{Kaplan:1998we,vanKolck:1998bw}. However it is a
convenient choice if no assumption is made about the relative size of
the renormalized EFT coefficients.

Now we should relate the scattering amplitude in the EFT, $A_2(p)$,
whose normalization is conventional and fixed to the Feynman diagram
expansion, to the S-matrix.  We will simply assume that the S-matrix
element for isotropic (s-wave) scattering exists in an arbitrary
number of spacetime dimensions. We then have generally
\beq
e^{2i\delta(p)}\ = \ 1\ +\ i\;{\cal N}(p)\;A_2(p) \ ,
\label{eq:3}
\eeq
where ${\cal N}(p)$ is a normalization factor that depends on $d$ and is fixed by unitarity.
Indeed combining eq.~(\ref{eq:2}) and eq.~(\ref{eq:3})
gives ${\cal N}(p)=-2{\rm Im}(I_0(p))$ and one can parametrize the scattering
amplitude by 
\beq
{\cal A}_2(p) \ = \ \frac{-1}{{\rm Im}(I_0(p))\big\lbrack\cot\delta(p)-i\big\rbrack} \ ,
\label{eq:4}
\eeq
with
\beq
\cot\delta(p) \ = \ \frac{1}{{\rm Im}(I_0(p))}\Bigg\lbrack\frac{1}{\sum C_{2n} \ p^{2n}}\ -\  {\rm Re}(I_0(p)) \Bigg\rbrack \ .
\label{eq:5}
\eeq
Bound states are present if there are poles on the positive imaginary momentum axis. That is if
$\cot\delta(i\gamma)=i$ with binding momentum $\gamma >0$.
These expressions are valid for any $d$. In order to evaluate $I_0(p)$ 
it is convenient to consider even and odd spacetime dimensions separately. 
For $d$ even the Gamma function has no poles and one finds 
\begin{eqnarray}
I_0(p)&=& -\frac{M}{2(4\pi)^{(d-1)/2}}\frac{\pi i\; p^{d-3}}{\Gamma\left({d-1\over 2}\right)}\ .
\label{eq:Ievend}
\end{eqnarray}
As there is no divergence, the $\overline{MS}$ EFT coefficients do not run with $\mu$ in
even spacetime dimensions. Hence the bare parameters are the renormalized parameters. For $d$ odd, one finds
\beq
I_0(p) =  \frac{M}{2(4\pi)^{(d-1)/2}}\frac{p^{d-3}}{\Gamma\left({d-1\over 2}\right)}
 \Bigg\lbrack \log{\left(-\frac{p^2}{\mu^2}\right)}\ -\ \psi_0\left({d-1\over 2}\right)\ - \log\pi\ -\ \frac{2}{\epsilon} \Bigg\rbrack \ ,
\label{eq:Iodd}
\eeq
where $\psi_0(n)$ is the digamma function. Here there is a single logarithmic divergence, hidden in the $1/\epsilon$ pole.
Hence in our scheme, at least one EFT coefficient runs with the scale $\mu$.
With these results in hand it is now straightforward to give the general expression for the
isotropic phase shift in $d$ spacetime dimensions:
\beq
p^{d-3}\cot\delta(p) \ = \  -\frac{(4\pi)^{(d-1)/2}}{\pi M}\Gamma\left({d-1\over 2}\right)\frac{2}{\sum C_{2n}\ p^{2n}}\ +\ (1-(-1)^d)\frac{p^{d-3}}{2\pi}\log{\left(\frac{p^2}{\overline{\mu}^2}\right)}\ ,
\label{eq:gencotdelta}
\eeq
where $\overline{\mu}$ is defined by equating the logarithm in eq.~(\ref{eq:gencotdelta}) with the content of the square
brackets in eq.~(\ref{eq:Iodd}). Note that this is an unrenormalized equation; the $C_{2n}$ coefficients are bare parameters
and there is a logarithmic divergence for odd spacetime dimensions. One must expand the right hand side of this equation for small momenta in order to 
renormalize~\footnote{In the case of three spatial dimensions eq.~(\ref{eq:gencotdelta}) yields the familiar effective range expansion,
\begin{eqnarray}
p\cot\delta(p) \ = \ -\frac{1}{a_{3}}\ +\ \frac{1}{2}r_{3}\, p^2 \ + \ {\mathcal O}( p^4 ) \ ,
\label{eq:7c}
\end{eqnarray}
with $a_{3} = {M C_0}/{(8\pi)}$ and $r_{3} = {16\pi C_2}/{(M C_0^2)}$.}.  It is noteworthy that
the effective field theory seems not to exist for $d>3$ and odd as the divergence is generated at leading order and
yet requires a nominally suppressed operator for renormalization.
%
\begin{figure}[t]
\includegraphics*[width=0.22\textwidth]{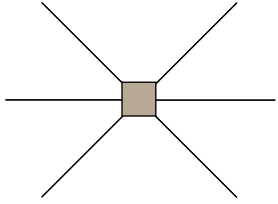}
\caption{\it Feynman diagram that gives the leading contribution to the three-body scattering amplitude.}
\label{fig:feynman2}
\end{figure}

The leading three-body diagram in the momentum expansion is shown in fig.~\ref{fig:feynman2}, and 
the three-body scattering amplitude is given by
\beq
{\cal A}_3 \ = \ -\ D_0 \ .
\label{eq:3bod1}
\eeq

\subsection{Two spatial dimensions}

\noindent In this section we consider the case $d=3$ in some detail. This case is particularly
interesting because of its analogy with renormalizable quantum field theories, and QCD in 
particular~\cite{Kaplan:2005es,Jackiw:1991je}.  From our general formula, eq.~(\ref{eq:gencotdelta}), we find
\begin{eqnarray}
\cot\delta(p) \ = \ \frac{1}{\pi} \log{\left(\frac{p^2}{\mu^2}\right)} \ -\ \frac{1}{\alpha_{2}(\mu)}\ +\ \sigma_2\, p^2 \ + \ {\mathcal O}( p^4 ) 
\label{eq:9}
\end{eqnarray}
where
\begin{eqnarray}
\alpha_{2}(\mu )\ =\ \frac{M C_0(\mu )}{8}\ ; \qquad  \sigma_2\ =\ \frac{8C_2(\mu)}{M C_0^2(\mu )} \ .
\label{eq:9b}
\end{eqnarray}
Note that $\alpha_2$ is a dimensionless coupling, and
$\sqrt{|\sigma_2|}$ is the effective range.  Neglecting range corrections, for $\alpha_{2}(\mu )$ of
either sign, there is a bound state with binding momentum $\gamma=\mu
\exp(\pi/2\alpha_{2}(\mu ))$. In essence, this occurs because,
regardless of the sign of the delta-function interaction, quantum
effects generate an attractive logarithmic contribution to the
effective potential which dominates at long distances. However, as we
will see below, in the repulsive case, this pole is not physical.

Many interesting properties in two spatial dimensions may be traced to scale invariance.
Keeping only the leading EFT operator, the Hamiltonian may be written as
\beq
{H}\ = \ \int d^2{\bf x}\Bigg\lbrack
{\textstyle \frac{1}{2}}\nabla{{\psi}^\dagger} \nabla \psi\ +\ 2 \alpha_2 ({{\psi}^\dagger} \psi)^2 \Bigg\rbrack\ ,
\label{eq:Ham}
\eeq where the field and spatial coordinates have been rescaled by
$\psi\rightarrow M^{1/2}\psi$; ${\vec x}\rightarrow M^{-1/2} {\vec
x}$. It is clear that classically there is no dimensionful parameter
and indeed this Hamiltonian has a non-relativistic conformal
invariance (Schr\"odinger invariance)~\cite{Jackiw:1991je}. This conformal invariance
is broken logarithmically by quantum effects.
Perhaps the most dramatic signature of this breaking of scale
invariance is the vanishing of the scattering amplitude at
zero energy, which follows from eqs.~\ref{eq:4} and \ref{eq:9}.

The leading beta function in the EFT is
\begin{eqnarray}
\mu\frac{d}{d\mu}C_0(\mu )\ = \ \frac{M}{4\pi}{C_0^2}(\mu )\ , 
\label{eq:10a}
\end{eqnarray}
which may be integrated to give the familiar renormalization group evolution equation
\begin{eqnarray}
\alpha_2(\mu )\ = \ \frac{\alpha_2(\nu)}{1- \frac{2}{\pi}\alpha_2(\nu )\log\left(\frac{\mu}{\nu}\right)} \ .
\label{eq:rg1}
\end{eqnarray}
It is clear from eq.~(\ref{eq:rg1}) that the attractive case,
$\alpha_2(\mu )=-|\alpha_2(\mu )|$, corresponds to an asymptotically
free coupling, while the repulsive case, $\alpha_2(\mu
)=+|\alpha_2(\mu )|$, has a Landau pole and the coupling grows weaker
in the infrared. We will focus largely on the latter case in what
follows~\footnote{For a recent discussion of the implications of scale
  invariance for many-boson systems in the case of an attractive
  coupling, see Ref.~\cite{Hammer:2004as}.}. Note that the position of the ``bound
state'' in the repulsive case coincides with the position of the
Landau pole, which sets the cutoff scale of the EFT. This state is
therefore unphysical.

Below we will also make use of a more conventional\footnote{With
  $a_2=a e^\gamma /2$ and $\sigma_2=a^2/2\pi$, this parametrization
  coincides with a hard-disk potential of radius $a$~\cite{schic}. As
  we will discuss below, there appears to be some confusion in the
  literature as regards the distinction between $a_2$ and $a$.}
parametrization of the phase shift:
\begin{eqnarray}
\cot\delta(p) \ = \ \frac{1}{\pi} \log{\left({p^2}{a_2^2}\right)} \ +\ \sigma_2\, p^2 \ + \ {\mathcal O}( p^4 ) \ .
\label{eq:9mod}
\end{eqnarray}
Here $a_2$ is the scattering length in two spatial dimensions. By matching with eq.~\ref{eq:9}, one
finds $a_2^{-1}=\mu \exp(\pi/2\alpha_{2}(\mu ))$, which in the repulsive case is the position of the
Landau pole. Hence, in the repulsive case, $a_2^{-1}$ is the momentum cutoff scale. Therefore,
from the point of view of the EFT, $a_2$ is a most unsuitable parameter for describing low-energy
physics. Of course, while the parameter $a_2$ is expected to be very small as compared to physical
scales, its effect is enhanced as it appears in the  argument of the logarithm.

\section{Scattering in a confined geometry}

\subsection{Eigenvalue equation}

\noindent With the scattering theory that we have developed we may now
find the eigenvalue equation in a confined geometry with periodic
boundary conditions. Specifically, we will consider scattering on what
is topologically the $(d-1)$-dimensional torus, ${\mathcal
T}^{d-1}={\mathcal S}_{(1)}^1\times{\mathcal
S}_{(2)}^1\times\cdots\times {\mathcal S}_{(d-1)}^1$.  In the confined
geometry, all bound and scattering states appear as poles of the
S-matrix, or scattering amplitude, $A_2(p)$.  Hence, from
eq.~(\ref{eq:2}) we have the eigenvalue equation $A_2(p)^{-1}=0$, or
\begin{eqnarray}
\frac{1}{\sum C_{2n} \ p^{2n}}\ =\  I^L_0(p) \qquad , \qquad I^L_0(p) \ =\ \frac{M}{2} \frac{1}{L^{d-1}}\sum^\Lambda_{\bf k} {1\over p^2-{\bf k}^2} \ ,
\label{eq:13}
\end{eqnarray}
where we have chosen to define the sum with a sharp cutoff ($d=2$ is ultraviolet finite).
The sum is over ${\bf k}=2\pi{\bf n}/L$ where
${\bf n}\in\mathbb{Z}^{d-1}=(n_1,n_2,\ldots,n_{d-1})$ takes all integer values. It is therefore convenient to write
\begin{eqnarray}
I^L_0(p) \ =\ \frac{M}{8\pi^2}{L^{3-d}}\sum^{\Lambda_n}_{{\bf n}\in\mathbb{Z}^{d-1}} {1\over q^2-{\bf n}^2} \ ,
\label{eq:15}
\end{eqnarray}
where $q\equiv {p L}/{2\pi}$ and therefore $\Lambda=2\pi\Lambda_n/L$. 
As the EFT coefficients are defined in DR, we can write the eigenvalue equation as
\begin{eqnarray}
\frac{1}{\sum C_{2n} \ p^{2n}}\ -\ {\rm Re}(I_0^{\lbrace DR\rbrace}(p))  \ =\  I^L_0(p) \ -\ {\rm Re}(I_0^{\lbrace \Lambda\rbrace}(p)) \ .
\label{eq:16}
\end{eqnarray}
Here we have subtracted off the real part of the loop integral using different schemes on the two sides
of the equation; the integral on the left is evaluated using DR and the one on the right is evaluated with a sharp cutoff
$\Lambda$. The purpose of this procedure is to leave the renormalization of the EFT coefficients, which is of course
an ultraviolet effect, unchanged while defining the integer sums using an integer cutoff.
We then have via eq.~(\ref{eq:5}) our general form for the eigenvalue equation
\begin{eqnarray}
\cot\delta(p) \ = \ \frac{1}{{\rm Im}(I_0(p))}\Big\lbrack I^L_0(p) \ -\ {\rm Re}(I_0^{\lbrace \Lambda\rbrace}(p)) \Big\rbrack \ .
\label{eq:17}
\end{eqnarray}

It is straightforward to find
\begin{eqnarray}
I_0^{\lbrace \Lambda\rbrace}(p)\ =\ \frac{M}{\left(4\pi\right)^{\frac{d-1}{2}}\Gamma\left(\frac{d-1}{2}\right)}
\frac{\Lambda^{d-1}}{(d-1)p^2}\  {}_2{\cal F}_1\left(1,\frac{d-1}{2},\frac{d+1}{2};\frac{\Lambda^2}{p^2}\right) \ ,
\label{eq:17b}
\end{eqnarray}
where ${}_2{\cal F}_1$ is the hypergeometric function.

The exact eigenvalue equation in $d$ spacetime dimensions can be written as
\begin{eqnarray}
\hspace{-0.17cm}q^{d-3}\cot\delta(p)\hspace{-0.12cm}
=\hspace{-0.12cm}\Gamma\left(\frac{d-1}{2}\right) \pi^{-\frac{d+1}{2}}\hspace{-0.21cm}\sum^{\Lambda_n}_{{\bf n}\in\mathbb{Z}^{d-1}} \frac{1}{{\bf n}^2-q^2} +   
\frac{2 \Lambda_n^{d-1}}{\pi (d-1)q^2} {\rm Re}\Bigg\lbrack{}_2{\cal F}_1\hspace{-0.13cm}\left(1,\frac{d-1}{2},\frac{d+1}{2};\frac{\Lambda_n^2}{q^2}\right)\hspace{-0.2cm}\Bigg\rbrack\hfill
\label{eq:18b}
\end{eqnarray}
where it is understood that $\Lambda_n\rightarrow\infty$ on the right
hand side. This equation gives the location of all of the
energy-eigenstates on the $(d-1)$-dimensional torus, including the
bound states (with $p^2 < 0$). The binding momentum in the confined
geometry reduces to $\gamma$ as $L\rightarrow\infty$.
While the derivation given above is valid within the radius of
convergence of the non-relativistic EFT, this eigenvalue equation is
expected to be valid for an arbitrary quantum field theory in $d$
dimensions up to corrections that are exponentially suppressed in the
boundary size, $L$.  One readily checks that eq.~\ref{eq:18b} gives the
familiar eigenvalue equations for $d=2$~\cite{Luscher:1990ck} and
$d=4$~\cite{Luscher:1986pf,Luscher:1990ux,Beane:2003da} and is in
agreement with Ref.~\cite{Fiebig:1994qi} for $d=3$.

\subsection{Two spatial dimensions}

\noindent In a finite area, the energy levels for the two-particle system follow from
the eigenvalue equation, eq.~(\ref{eq:18b}),
\begin{eqnarray}
\cot\delta(p) \ = \ \frac{1}{\pi^2}  \Bigg\lbrack {\cal S}_2\,\left( \frac{p L}{2 \pi} \right) \ +\ 
2\pi\log{\left(\frac{p L}{2 \pi} \right)} \Bigg\rbrack \ ,
\label{eq:24}
\end{eqnarray}
where
\begin{eqnarray}
{\cal S}_2\left( \eta \right)\ \equiv \  \sum^{\Lambda_n}_{\bf n} \frac{1}{ {\bf n}^2 - \eta^2} \ -\ 2\pi \log\Lambda_n \ .
\label{eq:25}
\end{eqnarray}
Using the results derived in Appendix II, this integer sum can be expressed as
\begin{eqnarray}
{\cal S}_2\left( \eta \right)\ =\ -\frac{1}{\eta^2}\ + {\cal P}_2 \ -\pi \gamma - 4\sum_{\ell=0}^\infty \frac{(-1)^\ell}{(2\ell +1)}\;\psi_0\left( 1 - \frac{\eta^2}{(2\ell +1)}\right) \ ,
\label{eq:25bintext}
\end{eqnarray}
where $\psi_0$ is the digamma function, and ${\cal P}_2$ is defined below. 

We can now combine our low-energy expansion, eq.~(\ref{eq:9}), with
the eigenvalue equation, eq.~(\ref{eq:24}), to find
\begin{eqnarray}
-\frac{1}{\alpha_2(\mu)}\ -\ \frac{2}{\pi} \log{\left(\frac{\mu L}{2\pi}\right)} \ +\ \sigma_2\, p^2 \ + \ {\mathcal O}( p^4 )  
\ = \ \frac{1}{\pi^2} {\cal S}_2\,\left( \frac{p L}{2 \pi} \right) \ .
\label{eq:26}
\end{eqnarray}
Using the renormalization group equation, eq.~\ref{eq:rg1}, we then have
\begin{eqnarray}
\cot\delta'(p) \ = \ \frac{1}{\pi^2} {\cal S}_2\,\left( \frac{p L}{2 \pi} \right) \ ,
\label{eq:27a}
\end{eqnarray}
where
\begin{eqnarray}
\cot\delta'(p) &\equiv & -\frac{1}{\alpha_2}\ +\ \sigma_2\, p^2 \ + \ {\mathcal O}( p^4 )  \ .
\label{eq:27}
\end{eqnarray}
and $\alpha_2\equiv\alpha_2({2\pi}/{L})$. We see that in the
eigenvalue equation, the logarithms of the energy cancel, and the
scale of the coupling is fixed to $2\pi/L$, the most infrared scale in the
EFT~\footnote{The prime on the phase shift indicates that the part of
  the scattering amplitude that is logarithmic in energy is
  removed. This is a consequence of the confined geometry.}. Therefore
as one approaches the continuum limit, the repulsive theory is at weak
coupling and the attractive theory is at strong coupling.

\subsection{Weak coupling expansion}

\noindent When the two-body interaction is repulsive, the eigenvalue equation, eq.~\ref{eq:27}, allows a weak coupling expansion of
the energy eigenvalues in the coupling $\alpha_2$. For the purpose of obtaining this expansion, it is convenient to rewrite the eigenvalue equation
in terms of the scale-invariant momentum ${\bf q}={\bf p}L/(2\pi)$. If one expands the energy in terms of the coupling one can
write $q^2=q_0^2+\epsilon q_1^2 +\epsilon^2 q_2^2+\ldots$, and the eigenvalue equation becomes
\begin{eqnarray}
-\frac{1}{\alpha_2}\ +\ \frac{\sigma_2\, (2\pi)^2}{L^2}\left( q_0^2 + \epsilon q_1^2  \ +\ \ldots \right)\ +\ \ldots 
&=& \epsilon\,\frac{1}{\pi^2} {\cal S}_2\,\left( q \right) \ .
\label{eq:wc1}
\end{eqnarray}
Note that in this expression, the range corrections break the scale
invariance with power law dependence on $L$. Indeed, in the presence
of the range corrections, one has a double expansion in $\alpha_2$ and
in $1/L^2$. It is now straightforward to compute the energy
perturbatively by expanding eq.~\ref{eq:wc1} in powers of $\epsilon$
and matching. 

With ${\bf q}_0=(0,0)$ one finds the ground-state energy
\begin{eqnarray}
E_0\ &=&\ \frac{4\alpha_2}{M L^2} \Bigg\lbrack\  1\ - \ \left(\frac{\alpha_2}{\pi^2}\right) {\cal P}_2\ +\ \left(\frac{\alpha_2}{\pi^2}\right)^2 \left( {\cal P}_2^2\ -\ {\cal P}_4\right)  
\ -\ \left(\frac{\alpha_2}{\pi^2}\right)^3 \left( {\cal P}_2^3\ -\ 3{\cal P}_2 {\cal P}_4\ +\ {\cal P}_6 \right) \nonumber \\
&& \qquad\qquad\qquad \ +\ {\cal O}(\alpha_2^4)\ \Bigg\rbrack \ +\ \frac{16\,\alpha_2^3\,\sigma_2}{M L^4}\left(1 \ +\ {\cal O}(\alpha_2) \right) 
\ +\ {\cal O}(L^{-6}) \ ,
\label{eq:27PT}
\end{eqnarray}
where we have included the leading range corrections and
\begin{eqnarray}
{\cal P}_2\ &\equiv& \ \sum^{\Lambda_n}_{{\bf n}\neq 0} \frac{1}{ {\bf n}^2} \ -\ 2\pi \log\Lambda_n \ =\  4\pi \log\left(e^{\frac{\gamma}{2}} \pi^{-\frac{1}{4}}\Gamma\left(\textstyle{\frac{3}{4}}\right)\right)\ =\ 2.5850  \ ; \ \nonumber \\ 
{\cal P}_4\ &\equiv& \ \sum^{\infty}_{{\bf n}\neq 0} \frac{1}{ {\bf n}^4} \ =\  \frac{2\pi^2}{3}{\cal C}\ =\ 6.0268   \ \ \  ; \ \ \
{\cal P}_6\ \equiv \ \sum^{\infty}_{{\bf n}\neq 0} \frac{1}{ {\bf n}^6} \ =\  \frac{\pi^3}{8}\zeta(3) \ =\ 4.6589 \ , \ 
\label{eq:27PT2}
\end{eqnarray}
where $\gamma$ is Euler's constant, ${\cal C}$ is Catalan's constant
and $\zeta(n)$ is the Riemann zeta function\footnote{These results are
  derived in Appendix I.}.  Note that one can use the renormalization
group to sum the terms of the form ${\cal O}({\alpha_2}^{n}{\cal P}_{2n})$. One finds, for instance,
for the universal part,
\begin{eqnarray}
E_0 &=& \frac{4\alpha_2'}{M L^2} \Bigg\lbrack\  1\ - \ \left(\frac{\alpha_2'}{\pi^2}\right)^2 {\cal P}_4  
\ -\ \left(\frac{\alpha_2'}{\pi^2}\right)^3 {\cal P}_6 \ +\ {\cal O}({\alpha_2'}^4)\ \Bigg\rbrack \ ,
\label{eq:27PT3}
\end{eqnarray}
where $\alpha_2'\equiv\alpha_2({2\pi}/{Lg})$ with $g\equiv e^{{\cal
    P}_2/(2\pi)}=1.5089$.  In this expression, the corrections to
leading order begin at ${\cal O}({\alpha_2'}^{3})$.  This freedom to
change the scale at which the coupling constant is evaluated will be
essential when we consider the many-body problem below.

\section{${\bf N}$ boson energy levels in a finite area}

\subsection{Perturbation theory for two identical bosons}

\noindent Exact eigenvalue equations for energy levels of ${\bf N}$
bosons (with ${\bf N}>2$) in a confined geometry are not available in
the EFT of contact operators in closed analytic form. Hence it is
worth asking whether the energy eigenvalues of the ${\bf N}$-body
problem admit sensible perturbative expansions about the free particle
energy. It is straightforward to approach this problem using
time-independent (Rayleigh-Schr\"odinger) perturbation theory. We will
first consider the two-body case. Consider the two-body
coordinate-space potential,
\begin{eqnarray}
V\left({\bf r}_1,{\bf r}_2\right)\ = \ \eta_2\;\delta^{d-1}\left({\bf r}_1-{\bf r}_2\right) \ ,
\label{eq:PT1}
\end{eqnarray}
where $\eta_2$ is the two-body pseudopotential, an energy-dependent function, which is determined by requiring that
the potential given by eq.~\ref{eq:PT1} reproduce the two-body scattering amplitude, eq.~\ref{eq:4}. 
The single-particle eigenfunctions in the confined geometry are
\begin{eqnarray}
\langle{\bf r}|{\bf p}\rangle\ =\ \frac{1}{L^{(d-1)/2}}\, e^{i{\bf p}\cdot{\bf r}} \ .
\label{eq:PT2}
\end{eqnarray}
The momentum-space two-body potential in the center-of-mass
system is then,
\begin{eqnarray}
V_{{\bf p},{\bf p}'}\ \equiv\ \langle{-{\bf p},{\bf p}}|V|{-{\bf p}',{\bf p}'}\rangle \ = \ \frac{\eta_2}{L^{d-1}} \ ,
\label{eq:PT3}
\end{eqnarray}
where $|{-{\bf p},{\bf p}}\rangle$ are the two-body unperturbed
eigenstates with energy $E_{\bf p}^0={\bf p}^2/M=(2\pi{\bf n}/L)^2/M$.
The perturbative expansion of the energy for momentum
level ${\bf n}$ is given by: 
\begin{eqnarray}
&&E_{\bf n} = \frac{4\pi^2{\bf n}^2}{ML^2} \ + \frac{\eta_2}{L^{d-1}} \ \Bigg\lbrack\ 1\ +\ 
\frac{\eta_2 M}{(2\pi)^2L^{d-3}} \sum_{{\bf m}\in\mathbb{Z}^{d-1}\neq {\bf n}}^{\Lambda_n} \frac{1}{{\bf n}^2-{\bf m}^2}  \nonumber \\
&&\hspace{-.45cm} + \left(\frac{\eta_2 M}{(2\pi)^2L^{d-3}}\right)^2\bigg\lbrack 
{\Bigl(\sum_{{\bf m}\in\mathbb{Z}^{d-1}\neq {\bf n}}^{\Lambda_n} \frac{1}{{\bf n}^2-{\bf m}^2}\Bigr)^2}-
\sum_{{\bf m}\in\mathbb{Z}^{d-1}\neq {\bf n}} \frac{1}{\left({\bf n}^2-{\bf m}^2\right)^2}\bigg\rbrack 
+ {\cal O}\Bigl(\Bigl(\frac{\eta_2}{L^{d-3}}\Bigr)^3\Bigr) \Bigg\rbrack .
\label{eq:PT4}
\end{eqnarray}{}
Hence for three spatial dimensions we have the nice perturbative
sequence $\lbrace 1/L^2,1/L^3,1/L^4,\ldots\rbrace$. However, in two
spatial dimensions we have $\lbrace 1/L^2,1/L^2,1/L^2,\ldots\rbrace$
with an energy independent two-body pseudopotential, and therefore
there is no perturbative expansion in $1/L$ about the free
energy, as expected on the basis of scale invariance. However, there
is, of course, an expansion in $\eta_2$ itself.

It is now straightforward to recover the perturbative expansion of the two-body ground
state energy in the case of two spatial dimensions. Here one finds
\begin{eqnarray}
\eta_2 \ =\  -{\frac{1}{2!}}\, {\cal A}^{\it tree}_2(p)\ =\
\frac{4\alpha_2}{M}\Bigl( 1\ +\ {\textstyle\frac{1}{2}}\,\sigma_2\alpha_2\left( {p^{\hspace{-.2cm}\leftarrow}}^2\ +\ {p^{\hspace{-.2cm}\rightarrow}}^2\right)\Bigr) \ ,
\label{eq:PT1a}
\end{eqnarray}
where the momenta have been written as arising from a Hermitian
operator.  In the relation between the pseudopotential and the
amplitude, the minus sign is from moving from the Lagrangian to the
Hamiltonian and $1/{\bf N}!$ is a combinatoric factor for ${\bf N}$
identical bosons. In order to deal
with the divergent sums in eq.~\ref{eq:PT4}, we renormalize as in the
exact case (eq.~\ref{eq:16}), and replace, for instance, the leading divergent
sum with
\begin{eqnarray}
\frac{M}{2(2\pi)^2L^{d-3}} \sum_{{\bf m}\in\mathbb{Z}^{d-1}\neq {\bf n}}^{\Lambda_n} \frac{1}{{\bf n}^2-{\bf m}^2} \ -\ 
{\rm Re}(I_0^{\lbrace \Lambda\rbrace}\left(p\right)) \ +\
{\rm Re}(I_0^{\lbrace DR\rbrace}\left(p\right)) \ .
\label{eq:PT5}
\end{eqnarray}
With $d=3$ and ${\bf n}=0$, this expression becomes
\begin{eqnarray}
-\frac{M}{2(2\pi)^2}\left( {\cal P}_2\ +\ 2\pi \log{\left(\frac{\mu L}{2\pi}\right)}\right) \ .
\label{eq:PT5a}
\end{eqnarray}
The scheme dependent part of the DR integral then defines the running coupling $\alpha_2(\mu)$.
Hence, inserting eq.~\ref{eq:PT1a} in eq.~\ref{eq:PT4}, and noting that 
the additional $\mu$-dependent piece in eq.~\ref{eq:PT5a} sets the scale of the coupling $\alpha_2$ 
to $2\pi/L$ as in the exact case considered above, one immediately recovers eq.~\ref{eq:27PT},
including the leading range corrections. We emphasize that the language of pseudopotentials
used here provides convenient bookkeeping in developing perturbation theory, however it is not 
essential.


\subsection{Perturbation theory for ${\bf N}$ identical bosons}

\noindent In this section, we generalize the perturbative expansion of
the ground-state energy to a system with ${\bf N}$ identical
bosons. The coordinate-space potential for the ${\bf N}$-body system
is
\begin{eqnarray}
V\left({\bf r}_1,\ldots ,{\bf r}_{\bf N}\right)\ = \ \eta_2\;\sum_{i<j}^{{\bf N}}\delta^{d-1}\left({\bf r}_i-{\bf r}_j\right) \ +\ 
\eta_3 \sum_{i< j<k}^{\bf N} \delta^{d-1}({\bf r}_i-{\bf r}_k)\delta^{d-1}({\bf r}_j-{\bf r}_k) +\ldots\,,
\label{eq:Nboson1}
\end{eqnarray}
where the dots denote higher-body operators. We have
\begin{eqnarray}
\eta_3 \ =\  -{\frac{1}{3!}} {\cal A}_3\ =\ {\frac{1}{6}} D_0 \ ,
\label{eq:Nboson1A}
\end{eqnarray}
where we have used eq.~\ref{eq:3bod1}.  It is straightforward but
unpleasant to find the ground-state energy of the ${\bf N}$ boson system
in perturbation theory with this potential.  In the case of three
spatial dimensions, this has been worked out up to order
$1/L^7$~\cite{Beane:2007qr,Tan:2007bg,Detmold:2008gh}. The calculation
in two spatial dimensions is essentially identical, as the combinatoric
factors for the ground-state level are independent of spatial
dimension, and therefore the dependence on spatial dimensionality resides
entirely in the coupling constant and the geometric constants.

\subsection{The ground-state energy}

\noindent In the case of two spatial dimensions one finds
the ground-state energy
\begin{eqnarray}
  E_0 &=& \frac{4\,\alpha_2}{M L^2}  \Bigg\lbrack\Choose{\bf N}{2}
  -\left(\frac{\alpha_2}{\pi^2}\right)\Choose{\bf N}{2} {\cal P}_2
+\left(\frac{\alpha_2}{\pi^2}\right)^2 \Bigg(
\Choose{\bf N}{2}{\cal P}_2^2
-\bigg\lbrack\Choose{\bf N}{2}^2 -12\Choose{\bf N}{3}-6\Choose{\bf N}{4}\bigg\rbrack{\cal P}_4
\Bigg)\nonumber
\\
&&\hspace*{-1.cm} +\ 
\left(\frac{\alpha_2}{\pi^2}\right)^3\Bigg(\hspace*{-.2cm}-\Choose{\bf N}{2} {\cal P}_2^3+3
\Choose{\bf N}{2}^2 {\cal P}_2{\cal P}_4 - \Choose{\bf N}{2}^3 {\cal P}_6 
-24 \Choose{\bf N}{3}\left( {\cal P}_2 {\cal P}_4+2
   {\cal Q}_0+{\cal R}_0- {\cal P}_6 \Choose{\bf N}{2}\right) 
\nonumber
\\&&
-6\Choose{\bf N}{4}\left(3 {\cal P}_2 {\cal P}_4+51 {\cal P}_6-2
   \Choose{\bf N}{2} {\cal P}_6\right) 
-300 \Choose{\bf N}{5}{\cal P}_6 -90
   \Choose{\bf N}{6}{\cal P}_6 \Bigg) \ +\ {\cal O}(\alpha_2^4) \Bigg\rbrack
\nonumber \\
& & \ +\ \frac{16\,\alpha_2^3\,\sigma_2}{M L^4}{{\bf N} \choose 2} \ +\ \frac{1}{L^4}\frac{D_0}{6}{{\bf N} \choose 3} \ ,
  \label{eq:GSEwBC}
\end{eqnarray}
where ${\tiny \Choose{n}{k}}$=$n!/(n-k)!/k!$, the ${\cal P}_{2s}$ are available 
in eq.~\ref{eq:27PT2}, and
\begin{eqnarray}
{\cal Q}_0 &=& 
\sum_{\bf n\ne 0}\sum_{\bf m\ne 0} 
\frac{1}{{\bf  n}^2\ {\bf m}^2\ 
({\bf n}^2 + {\bf m}^2 + ({\bf n}+{\bf m})^2)}\ =\ 16.3059 \ ;
\,   \label{eq:10}
\\
{\cal R}_0 &=& 
\sum_{\bf m\ne 0} \sum_{\bf n}^{\Lambda_n}\ {1\over {{\bf m}^4(\bf n}^2 + {\bf m}^2 + ({\bf n}+{\bf m})^2)}
\ -\ \pi {\cal P}_4\,\log\Lambda_n \ =\  -1.8414 \ .   
\label{eq:11}
\end{eqnarray}
These double lattice sums have been evaluated numerically. 
This expression for the ground-state energy is complete to ${\cal  O}(\alpha_2^4)$, and includes
the leading non-universal effects due to range corrections and three-body forces. Expanding out the binomial
coefficients gives, for the universal piece,
\begin{eqnarray}
&& E_0\ =\ \frac{4\,\alpha_2}{M L^2}{{\bf N} \choose 2}   \Bigg\lbrack\  1\ - \ \left(\frac{\alpha_2}{\pi^2}\right) {\cal P}_2\ +\ \left(\frac{\alpha_2}{\pi^2}\right)^2 \Bigl( {\cal P}_2^2\ +\ (2{\bf N}-5) {\cal P}_4\Bigr)  \nonumber \\
& & \ -\ \left(\frac{\alpha_2}{\pi^2}\right)^3 \Bigl( {\cal P}_2^3\ +\ (2{\bf N}-7){\cal P}_2 {\cal P}_4\ +\ (5{\bf N}^2 - 41 {\bf N} +63) {\cal P}_6 
\ +\ 8({\bf N}-2)(2{\cal Q}_0\ +\ {\cal R}_0)\Bigr)
\nonumber \\
& & \ +\ \left(\frac{\alpha_2}{\pi^2}\right)^4 \Bigl( {\cal P}_2^4\ 
-\ 6 {\cal P}^2_2 {\cal P}_4 
\ +\ (4+{\bf N}-{\bf N}^2){\cal P}^2_4 
\ +\ 4(27-15{\bf N}+{\bf N}^2){\cal P}_2 {\cal P}_6  \Bigr. \nonumber \\
& &\Bigl. \qquad \qquad\qquad\ +\ (14{\bf N}^3-227{\bf N}^2+919{\bf N}-1043){\cal P}_8\ +\ \ldots\
\Bigr) \ +\ {\cal O}(\alpha_2^5) \Bigg\rbrack \ .
\label{eq:Nboson3}
\end{eqnarray}
Here we have included a part of the ${\cal O}(\alpha_2^5)$ 
contribution~\cite{Detmold:2008gh} for reasons that will become clear in the
next section. The dots represent double and triple lattice sums
that appear at ${\cal O}(\alpha_2^5)$~\cite{Detmold:2008gh,Detmold:UN} and which
we do not consider here.

As $\alpha_2\equiv\alpha_2({2\pi}/{L})$, it is clear that this expansion is valid
only for repulsive coupling, which is small in the infrared. The expansion is 
expected to be valid for $|\alpha_2| {\bf N} \ll 1$~\footnote{For a more accurate measure of the
regime of applicability of the expansion, see Appendix II.}. 
The chemical potential and pressure are readily available from the ground-state
energy via the formulas
\begin{eqnarray}
\mu \ =\ \frac{d E_0}{d{\bf N}}\Big|_L \qquad ;  \qquad {\cal P} \ =\ -\frac{1}{2L}\frac{d E_0}{dL}\Big|_{\bf N} \ .
\label{eq:thermo}
\end{eqnarray}

By inspection of the binomial coefficients in eq.~\ref{eq:GSEwBC} one
sees that the leading effects from three-body forces enter at ${\cal
  O}(\alpha_2^3)$ and ${\cal O}(\alpha_2^4)$ through two-body
interactions. Other three-body effects enter by way of the contact
operator in the Lagrangian and appear at the same order as effective
range corrections: that is, they are suppressed by $L^{-2}$ as
compared to the leading two-body contributions, treated to all orders.
This is of course a consequence of scale invariance. It is worth
noting the contrast with the case of three spatial
dimensions~\cite{Beane:2007qr,Tan:2007bg,Detmold:2008gh}.  There, the
two-body contributions to the three-body force are logarithmically
divergent in the ultraviolet and are renormalized by the three-body
force contact operator in the Lagrangian. Both effects appear at
${\cal O}(L^{-6})$ in the expansion of the ground-state energy.  In
two spatial dimensions, scale invariance ensures that the sums in
eqs.~\ref{eq:10} and \ref{eq:11} are convergent as there is no
scale-invariant counterterm available. Moreover, this ultraviolet
finiteness persists to arbitrary order in $\alpha_2$.

\section{The thermodynamic limit and the density expansion}

\subsection{The Lee-Huang-Yang strategy}

\noindent The underlying scale invariance of the two-dimensional
system allows a great deal to be learned about the thermodynamic limit
directly from from the expression for the ${\bf N}$-body ground-state
energy in a finite area.  By thermodynamic limit we intend the limit
where ${\bf N}$ and $L$ are taken to infinity with the density,
$\rho\equiv{\bf N}/L^2$, held fixed. Our strategy will be as
follows. First we will use the renormalization group equation for the
coupling $\alpha_2$ to change the scale at which the coupling is
evaluated to a quantity that is finite in the thermodynamic limit. We
will then rearrange the expansion of the energy into series determined
according to the degree of divergence with large ${\bf
  N}$~\cite{Lee:1957zzb}. These series must sum to finite quantities
in the thermodynamic limit. All quantities that are finite in this
limit are kept. We will see that this strategy will enable us to
constrain the form of the low-density expansion of the energy density
of the two-dimensional Bose gas. Moreover, we will see that, unlike in
the case of three spatial dimensions, the series that are most divergent
with ${\bf N}$ can be explicitly evaluated.

\subsection{Universality and broken scale invariance}

\noindent As the coupling $\alpha_2$ in eq.~\ref{eq:Nboson3} is
evaluated at the far infrared scale $2\pi/L$, a change of scale is
required before performing the thermodynamic limit. Consider a change
of scale to $\nu = 2\pi \sqrt{\lambda\rho}$, where $\lambda$ is a
number which represents the inherent ambiguity in the choice of
scale. With this choice, ${\alpha_2}$ is finite in the thermodynamic
limit, and constitutes a small parameter in the low-density limit
(assuming that $\lambda$ is a number of order unity.) Using
eq.~\ref{eq:rg1}, we can then reexpress the energy as
\begin{eqnarray}
E_0\ &=&\ \frac{4\alpha_2'}{M L^2}{{\bf N} \choose 2}   \Bigg\lbrack\  1\ 
- \ \left(\frac{\alpha_2'}{\pi^2}\right) \Bigl( {\cal P}_2  \ +\ \pi\log\left({\bf N}\lambda\right)\Bigr)\nonumber \\
& &\ +\ \left(\frac{\alpha_2'}{\pi^2}\right)^2 \Bigl( {\cal P}_2^2\ +\ (2{\bf N}-5) {\cal P}_4 \ +\ \pi\log\left({\bf N}\lambda\right)\Bigl(2 {\cal P}_2 \ +\ 
\pi\log\left({\bf N}\lambda\right)   \Bigr)            \Bigr)  \nonumber \\
& & \ -\ \left(\frac{\alpha_2'}{\pi^2}\right)^3 \Bigl( {\cal P}_2^3\ +\ (2{\bf N}-7){\cal P}_2 {\cal P}_4\ +\ (5{\bf N}^2 - 41 {\bf N} +63) {\cal P}_6 
\ +\ 8({\bf N}-2)(2{\cal Q}_0\ +\ {\cal R}_0)\Bigr.
\nonumber \\ 
& & \Bigl. \quad\qquad\ +\ \pi\log\left({\bf N}\lambda\right)\left( 3\left( {\cal P}_2^2 \ +\ \left(2{\bf N} - 5\right){\cal P}_4\right)\ +\  \pi\log\left({\bf N}\lambda\right)
\Bigl( 3 {\cal P}_2 \ + \ \pi\log\left({\bf N}\lambda\right) \Bigr)\right)\Bigr)
\nonumber \\ 
& & \ +\ \left(\frac{\alpha_2'}{\pi^2}\right)^4 \left( {\cal P}_2^4\ 
-\ 6 {\cal P}^2_2 {\cal P}_4 
\ +\ (4+{\bf N}-{\bf N}^2){\cal P}^2_4 
\ +\ 4(27-15{\bf N}+{\bf N}^2){\cal P}_2 {\cal P}_6  \right. \nonumber \\
& &\left. \quad\qquad \ +\ (14{\bf N}^3-227{\bf N}^2+919{\bf N}-1043){\cal P}_8 \ +\ \ldots\
\right)
\ +\ {\cal O}({\alpha_2'}^5) \Bigg\rbrack \ ,
\label{eq:Nboson4}
\end{eqnarray}
where now $\alpha_2'\equiv\alpha_2(\nu)$. This expression is
independent of $\lambda$  up to ${\cal O}({\alpha_2'}^5)$ corrections. The strategy is to rearrange the expansion
according to the maximum powers of ${\bf N}$ that appear at each order in ${\alpha_2'}$.
We can then re-write eq.~\ref{eq:Nboson4} as the energy-per-particle:
\begin{eqnarray}
\frac{E_0}{\bf N}\ &=&\ \frac{2\alpha_2'}{M}\left( \rho \ + \ \frac{1}{L^2}\right)  \Bigg\lbrack\  1\ + \ 
\frac{1}{\bf N}\;{\cal G}\ +\ 
\frac{1}{{\bf N}^2}\;\Bigl(\pi\log\left({\bf N}\lambda\right){\cal H}+{\cal I}\Bigr)
\nonumber \\ 
&&\ - \ \left(\frac{\alpha_2'}{\pi^2}\right) \Bigl( {\cal P}_2 \ +\ \pi\log\left({\bf N}\lambda\right)\Bigr)\nonumber \\
& &\ +\ \left(\frac{\alpha_2'}{\pi^2}\right)^2 \Bigl( {\cal P}_2^2\ -\ 5 {\cal P}_4 \ +\ \pi\log\left({\bf N}\lambda\right)\Bigl(2 {\cal P}_2 \ +\ 
\pi\log\left({\bf N}\lambda\right)   \Bigr)         \Bigr) \ +\ {\cal O}({\alpha_2'}^3)  \Bigg\rbrack
\label{eq:Nboson5}
\end{eqnarray}
where
\begin{eqnarray}
{\cal G}\left(z \right)&=&2\,z^2\,  {\cal P}_4\ -\  5\,z^3\,{\cal P}_6 \ +\ 14\,z^4\,{\cal P}_8 \ +\ {\cal O}(z^5) \label{eq:Gdefined2} \\
{\cal H}\left(z \right)&=&-6\, z^3\,  {\cal P}_4\ +\ 20\, z^4\,{\cal P}_6\ -\ 70\, z^5\, {\cal P}_8 \ +\ {\cal O}(z^6) \label{eq:Hdefined2} \\
{\cal I}\left(z \right)&=&\, -z^3 \Bigl(\ 2{\cal P}_2{\cal P}_4\ -\ 41 {\cal P}_6 \ +\ 8(2{\cal Q}_0\ +\ {\cal R}_0)\ \Bigr) \label{eq:Idefined2}\\
&&\qquad\qquad\qquad  +\ z^4\left( 4{\cal P}_4{\cal P}_6\ -\ {\cal P}_4^2\ +\  227{\cal P}_8\ +\ \ldots   \right)  \ +\ {\cal O}(z^5) \ ,
\end{eqnarray}
with $z\equiv{\bf N}\alpha_2'/{\pi^2}$. The mathematically-inclined
reader will immediately notice that the coefficients of the first two
sums are related to the Catalan numbers. We will postpone till later
discussion of the evaluation of these sums, in order to focus on
obtaining the form of the low-density expansion which is based purely on general
considerations. It is clear from eq.~\ref{eq:Nboson5} that in order to
have a finite thermodynamic limit, ${\cal G}(z)$ must scale as $z$ and
${\cal H}(z)$ and ${\cal I}(z)$ must scale as $z^2$ for large
$z$. Hence we may define
\begin{eqnarray}
\lim_{z\to\infty}\frac{1}{z}\,{\cal G}(z)\ \equiv\ {\mathfrak g}(z)\ ; \ \ \lim_{z\to\infty}\frac{1}{z^2}\,{\cal H}(z)\ \equiv\ {\mathfrak h}(z)\ ; \ \ 
\lim_{z\to\infty}\frac{1}{z^2}\,{\cal I}(z)\ \equiv\ {\mathfrak i}(z) \ ,
\label{eq:Gdefined3}
\end{eqnarray}
where ${\mathfrak g}(z)$, ${\mathfrak h}(z)$ and ${\mathfrak i}(z)$ have, at most, logarithmic dependence on $z$.
In the limit that ${\bf N}$ and $L$ are large but finite we then have:
\begin{eqnarray}
\frac{E_0}{\bf N}&&=\ \frac{2\alpha_2'\rho}{M} \Bigg\lbrack\  1\ + \ 
\left(\frac{\alpha_2'}{\pi^2}\right) \Bigl( {\mathfrak g}\ -\ {\cal P}_2 \ -\ \pi\log\left({\bf N}\lambda\right)\Bigr)\nonumber \\
& &\hspace*{-1.cm} + \left(\frac{\alpha_2'}{\pi^2}\right)^2 \Bigl( {\mathfrak i}\ +\ {\cal P}_2^2\ -\ 5 {\cal P}_4 \ +\ 
\pi\log\left({\bf N}\lambda\right)\Bigl( {\mathfrak h}\ +\ 2 {\cal P}_2 \ +\  \pi\log\left({\bf N}\lambda\right)   \Bigr)\Bigr) \ +\ {\cal O}({\alpha_2'}^3) 
\Bigg\rbrack.
\label{eq:Nboson5A}
\end{eqnarray}
This form makes clear that the logarithmic dependence on ${\bf N}$ must be canceled by
${\mathfrak g}(z)$, ${\mathfrak h}(z)$ and ${\mathfrak i}(z)$ in order to be left with an energy-per-particle 
that is finite in the thermodynamic limit. That is, we have the differential equations,
\begin{eqnarray}
&&\frac{d}{d{\bf N}}\Bigl( {\mathfrak g}\ -\ {\cal P}_2 \ -\ \pi\log\left({\bf N}\lambda\right)\Bigr)\ =\ 0 \ ;\nonumber \\
&&\frac{d}{d{\bf N}}\Bigl( {\mathfrak i}\ +\ {\cal P}_2^2\ -\ 5 {\cal P}_4 \ +\ 
\pi\log\left({\bf N}\lambda\right)\Bigl( {\mathfrak h}\ +\ 2 {\cal P}_2 \ +\  \pi\log\left({\bf N}\lambda\right) \Bigr) \Bigr) \ =\ 0 \ ,
\label{eq:Nboson5B}
\end{eqnarray}
which are readily integrated to give:
\begin{eqnarray}
{\mathfrak g}(z) &=&  \pi \log z\ +\ {\bar g}\ ;\label{eq:gdefined4} \\
{\mathfrak h}(z) &=&   -2\pi \log z\ +\ {\bar h}\ ; \label{eq:hdefined4} \\
{\mathfrak i}(z) &=&  \pi^2 \log^2 z\ -\ \pi\left( {\bar h}\ +\ 2{\cal P}_2\right)\,\log z \ +\ {\bar i} \label{eq:idefined4} \ ,
\end{eqnarray}
where ${\bar g}$, ${\bar h}$ and ${\bar i}$ are integration constants. Plugging these functions into eq.~\ref{eq:Nboson5A} we may take the
thermodynamic limit and we find to ${\cal O}({\alpha_2'}^3)$:
\begin{eqnarray}
\frac{E_0}{\bf N}&=& \frac{2\alpha_2'\rho}{M} \Bigg\lbrack\  1\ + \ 
\left(\frac{\alpha_2'}{\pi}\right) \Bigl( \log\alpha_2' \ -\ \log\lambda\pi^2 \ -\ \textstyle{\frac{1}{\pi}}\left({\cal P}_2\ -\ {\bar g}\right) \Bigr) \nonumber \\ 
&& +\ \left(\frac{\alpha_2'}{\pi}\right)^2 \Bigl( \log^2\alpha_2'\ -\ \left( 2\log\lambda\pi^2\ +\  
\textstyle{\frac{1}{\pi}}\left(2{\cal P}_2\ +\ {\bar h}\right)
\right)\log\alpha_2'   \Bigr. \nonumber \\
&& \Bigl.  +\  \log\lambda\pi^2\, \textstyle{\frac{1}{\pi}}\left(2{\cal P}_2\ +\ {\bar h}\right) \ +\  \log^2\lambda\pi^2 
\ +\ \frac{1}{\pi^2}\left( {\cal P}_2^2\ -\ 5 {\cal P}_4 \ +\ {\bar i}\right)\Bigr)  \nonumber \\
&& +\ {\cal O}\left({\alpha_2'}^3\right)
\Bigg\rbrack \ .
\label{eq:NbosonwithsumsB}
\end{eqnarray}
There is one further constraint: here we expect that the energy-per-particle should be independent of $\lambda$ up to 
${\cal O}({\alpha_2'}^{4})$ corrections. Using eq.~\ref{eq:rg1}, one finds
\begin{eqnarray}
\lambda \frac{d}{d\lambda}\left(\frac{E_0}{\bf N}\right)&=& \frac{2{\alpha_2'}^3\rho}{M\pi^3} \left(2{\bar g}\ +\ {\bar h}\ +\ \pi\right)\ +\ 
{\cal O}({\alpha_2'}^{4}) \ ,
\label{eq:RGconstraing}
\end{eqnarray}
and therefore there is a further relation between the integration constants, 
\begin{eqnarray}
2{\bar g} + {\bar h} + \pi=0 \ .
\label{eq:RGconstraingB}
\end{eqnarray}
Note that in eq.~\ref{eq:NbosonwithsumsB}, the energy is completely
determined to ${\cal O}({\alpha_2'}^2\log\alpha_2')$.  Indeed we see
that the coefficients of the leading logarithms of the form ${\cal
  O}({\alpha_2'}^{n+1}\log^n\alpha_2')$ are fixed. 
The change of renormalization scale to obtain a
density-dependent coupling introduced terms of the form $\log {\bf N}$,
and as the form of the density expansion had to be such as to cancel
these divergent terms, it is not surprising that the leading logarithms 
in the expansion can be removed by a change of scale. (We will
do this explicitly below.)

In order to go further, one must evaluate the sums,
eqs.~\ref{eq:Gdefined2}, \ref{eq:Hdefined2} and \ref{eq:Idefined2}. We
evaluate ${\cal G}$ and ${\cal H}$ in Appendix II. We recover the form
as expected in eq.~\ref{eq:idefined4} and find the integration
constants to be:
\begin{eqnarray}
{\bar g}&=&  \frac{\pi}{2}\ +\ {\cal P}_2 \ ;\label{eq:gbardefined4} \\
{\bar h}&=&  -2\pi\ -\ 2{\cal P}_2 \ ,\label{eq:ibardefined4} 
\end{eqnarray}
which are of course consistent with eq.~\ref{eq:RGconstraingB}.
We have been unable to determine ${\bar
  i}$~\footnote{By inspection of eq.~\ref{eq:Hdefined2}, it would
  appear that performing the sum to obtain ${\cal H}$ would involve
  solving the two-dimensional double lattice sums ${\cal Q}_0$ and
  ${\cal R}_0$ in the sense of expressing them as products of
  one-dimensional sums.}. The two integration constants ${\bar g}$ and
${\bar h}$ then fix the energy density to ${\cal O}({\alpha_2'}^3\log\alpha_2')$:
\begin{eqnarray}
&&{\cal E}\left(\rho\right) = \rho\times \frac{E_0}{\bf N}=\ \frac{2\alpha_2'\rho^2}{M} \Bigg\lbrack\  1\ + \ 
\left(\frac{\alpha_2'}{\pi}\right) \Bigl( \log\alpha_2' \ -\ \log\lambda\pi^2 \ +\ \textstyle{\frac{1}{2}} \Bigr) \nonumber \\ 
&& + \left(\frac{\alpha_2'}{\pi}\right)^2 \Bigl( \log^2\alpha_2' + 2(1 - \log\lambda\pi^2)\log\alpha_2' 
 + \log\lambda\pi^2 \left(\log\lambda\pi^2\ -\ 2\right)
 + {\textstyle \frac{1}{\pi^2}}\left( {\cal P}_2^2 - 5 {\cal P}_4  + {\bar i}\right)\Bigr)\nonumber \\
&&\qquad\qquad\qquad  +\ {\cal O}\left( {\alpha_2'}^3\right) 
\Bigg\rbrack \ .
\label{eq:Nbosonwithsums}
\end{eqnarray}
It is straightforward to check that this result is in 
agreement with Refs.~\cite{chern} and \cite{Andersen} to ${\cal
  O}({\alpha_2'}^3\log\alpha_2')$. As the energy density in the thermodynamic limit
cannot depend on the geometric constants ${\cal P}_2$ and ${\cal P}_4$, we define
\begin{eqnarray}
C \equiv -\frac{1}{2}\Bigl( 1 \ +\ 
\frac{1}{\pi^2}\left( {\cal P}_2^2 - 5 {\cal P}_4  + {\bar i}\right)\Bigr) \ .
\label{eq:AnderTRANS}
\end{eqnarray}
While we have been unable to calculate this constant, Ref.~\cite{Andersen} finds
\begin{eqnarray}
C_A &=& 2.78 \times 10^{-3}\ .
\label{eq:AnderTRANS2}
\end{eqnarray}
In the calculation of Ref.~\cite{Andersen}, which is based on a
systematic EFT computation of quantum fluctuations around a mean
field~\cite{braa}, $C$ arises as a two-loop effect. This is consistent
with our expectations for the integration constant ${\bar i}$, as the
leading term in the sum depends on the double lattice sums ${\cal
  Q}_0$ and ${\cal R}_0$ which are clearly related to two-loop vacuum
integrals in the continuum limit.

\subsection{Finite-size corrections}

\noindent As we are able to evaluate the sums ${\cal G}$ and ${\cal H}$ explicitly, we are able
to give the leading and next-to-leading finite-size corrections to the thermodynamic limit. Relaxing
the thermodynamic limit in eq.~\ref{eq:Gdefined3} gives:
\begin{eqnarray}
\frac{1}{z}\,{\cal G}(z)\ \equiv\ {\mathfrak g}(z) \ +\ \frac{{\mathfrak g}_0}{z}\ +\ {\cal O}(z^{-2})
\ ; \ \ 
\frac{1}{z^2}\,{\cal H}(z)\ \equiv\ {\mathfrak h}(z) \ +\ \frac{{\mathfrak h}_0}{z}\ +\ {\cal O}(z^{-2}) \ .
\label{eq:Gdefined3FS}
\end{eqnarray}
In Appendix II we find ${\mathfrak g}_0 = 1 = -{\mathfrak h}_0$.  Using eq.~\ref{eq:Nboson5} we see that
there are leading and subleading $1/{\bf N}$ corrections that arise from  ${\mathfrak g}_0,{\mathfrak h}_0\neq 0$,
and, in addition, there is a correction that arises from the binomial coefficient prefactor, as shown explicitly in 
eq.~\ref{eq:Nboson5}. Taking into account both of these contributions gives
\begin{eqnarray}
\delta{\cal E}_{FS} \ =\  
\frac{4\alpha_2'\rho^2}{M{\bf N}} \Bigg\lbrack\  1\ + \ \left(\frac{\alpha_2'}{2\pi}\right) \Bigl( \log\alpha_2' \ -\ \log{\bf N}\lambda^2\pi^2 \ +\ \textstyle{\frac{1}{2}} \Bigr) \Bigg\rbrack \ .
\label{eq:NbosonwithsumsFS}
\end{eqnarray}
One easily checks that this expression is independent of $\lambda$  up to ${\cal O}({\alpha_2'}^3)$ corrections. It follows
that
\begin{eqnarray}
\frac{\delta{\cal E}_{FS}}{\cal E} \ =\  \frac{2}{\bf N} \ +\ {\cal O}({\alpha_2'}) \ ,
\label{eq:NbosonwithsumsFS2}
\end{eqnarray}
which constitutes a ten-percent effect in a system with ${\bf N}=20$ bosons.

\subsection{Non-universal corrections}

\noindent It is straightforward to include the leading non-universal corrections in
the energy density. By inspection of eq.~\ref{eq:Nboson3}, it is clear that the
leading effective range corrections to the ground-state energy of ${\bf N}$ bosons in a finite
area vanishes in the thermodynamic limit. Hence, the leading non-universal contribution
to the energy density is from the three-body force,
\begin{eqnarray}
\delta{\cal E}_{NU}\ =\ \rho^3\, \frac{D_0}{36} \ ,
\label{eq:NbosonwithsumsNU}
\end{eqnarray}
as one expects on the basis of simple mean field theory considerations.
An estimate of the leading range corrections to the energy density has been made in Ref.~\cite{astrNU}.

\subsection{Summary and discussion}

\noindent Our final form for the energy density in the thermodynamic limit may be written as
\begin{eqnarray}
&& {\cal E}\ +\ \delta{\cal E}_{NU} \ =\ \frac{2\alpha_2'\rho^2}{M} \Bigg\lbrack\  1\ + \ 
\left(\frac{\alpha_2'}{\pi}\right) \Bigl( \log\alpha_2' \ -\ \log\lambda\pi^2 \ +\ \textstyle{\frac{1}{2}} \Bigr) \nonumber \\ 
&&+\ \left(\frac{\alpha_2'}{\pi}\right)^2 \Bigl( \log^2\alpha_2'\ +\ 2(1\ -\ \log\lambda\pi^2)\log\alpha_2' 
\ +\ \log\lambda\pi^2 \left(\log\lambda\pi^2\ -\ 2\right)\ -\ 1\ -\ 2 C  \Bigr)\nonumber \\
&& +\ {\cal O}\left( {\alpha_2'}^3\right) 
\Bigg\rbrack \ +\ \rho^3\, \frac{D_0}{36} \ ,
\label{eq:NbosonwithsumsSUMM}
\end{eqnarray}
which is complete to ${\cal O}({\alpha_2'}^3)$ if one uses
eq.~\ref{eq:AnderTRANS2} for the constant $C$ from Ref.~\cite{Andersen}, and is
valid for $\alpha_2'\log\alpha_2'\ll 1$.  We stress that
this result is model independent; the critical assumption that we have
made is that the potentials via which the bosons interact are of
finite range.  The chemical potential and pressure are readily
available from the energy density via
\begin{eqnarray}
\mu \ =\ \frac{d{\cal E} }{d{\rho}} \qquad ;  \qquad {\cal P} \ =\ -{\cal E}\ +\ \rho \mu \ .
\label{eq:thermo2}
\end{eqnarray}

The results found in this paper (and also in Ref.~\cite{Andersen})
demonstrate that there is a scale ambiguity in the equation-of-state
of the Bose gas in two spatial dimensions. While the energy density is
independent of $\lambda$ and of the renormalization scheme that is
used to define the coupling constants, in the perturbative expansion
this holds only up to ${\cal O}({\alpha_2'}^{4})$ corrections; that
is, the inevitable truncation of the perturbative expansion implies
that predictions do depend on $\lambda$. In principle, an ideal choice
for $\lambda$ will optimize perturbation theory for the particular
system in question~\footnote{Similar scale ambiguities arise in
  perturbative QCD. For a relevant discussion, see
  Refs.~\cite{Brodsky:1982gc} and \cite{Brodsky:1994eh}.}.  For
instance, if one chooses $\lambda=\alpha_2'/\pi^2$, then all
logarithms of the coupling are absorbed into the coupling itself and
one is left with a simple perturbative expansion in $\alpha_2'$,
\begin{eqnarray}
&& {\cal E} \ =\ \frac{2\alpha_2'\rho^2}{M} \Bigg\lbrack\  1\ + \ 
\left(\frac{\alpha_2'}{\pi}\right) \Bigl( {\frac{1}{2}} \Bigr) \ -\ \left(\frac{\alpha_2'}{\pi}\right)^2 \Bigl( 1\ +\ 2 C  \Bigr)
\ +\ {\cal O}\left( {\alpha_2'}^3\right)  \Bigg\rbrack \ ,
\label{eq:weirdscale}
\end{eqnarray}
where the coupling $\alpha_2'$ is determined self-consistently from eq.~\ref{eq:rg1}
or from eq.~\ref{eq:schick} below.

The most natural way of expressing interactions in the EFT is in terms
of the Lagrangian coefficients, which run with the renormalization
group in two spatial dimensions. By contrast, the two-dimensional
scattering length is not a natural quantity in the EFT; indeed it is
the most unnatural quantity that it is possible to form, as it
corresponds to the distance scale set by the Landau pole.
Nevertheless, the energy density can be expressed in terms of the
two-dimensional scattering length via the formula
\begin{eqnarray}
\alpha_2'\ &=& -\frac{\pi}{\log{(\rho\lambda(2\pi)^2 a_2^2)}} \ ,
\label{eq:schick}
\end{eqnarray}
which is obtained by comparing eq.~\ref{eq:9} and eq.~\ref{eq:9mod}.
This is the traditional way of expressing the two-body coupling
constant~\cite{schic}. We see that the argument of the logarithm
depends on $\lambda$, and is therefore not a physical quantity; any
attempt to assign definite meaning to it is futile.

Finally, for facility in comparison, we will express the universal
part of the energy density in terms of the scattering length.  As
pointed out in section II, there are various conventions used in the
literature for the scattering length; one convention, $a_2$, is as
given in eq.~\ref{eq:9mod} and another identifies the scattering
length with the radius of a hard disc, $a$~\footnote{Evidently
  Refs.~\cite{astrDC,moraDC,astrNU} claim that Refs.~\cite{chern} and
  \cite{Andersen} are discrepant, and, moreover, that
  Ref.~\cite{Andersen} is incorrect. As pointed out above, we find no
  discrepancy between these two calculations.  We believe that
  confusion may have arisen due to the choice of convention for the
  two-dimensional scattering length.}.  In the first convention, we
have, with $\lambda=1/(2\pi)^2$,
\begin{eqnarray}
{\cal E}& =& \frac{2\pi\rho^2}{M |\log\rho{ a_2^2}|}\Bigg\lbrack\  
1\ - \ \frac{1}{|\log\rho{ a_2^2}|} \Bigl( \log |\log\rho {a_2^2}|\ -\ \log4\pi \ -\ {\textstyle \frac{1}{2}}  \Bigr) \nonumber \\
&&\qquad\qquad\qquad\qquad +\ \frac{1}{|\log\rho{ a_2^2}|^2}\Bigl(\log^2 |\log\rho {a_2^2}|\ -\ 2(1\ +\ \log4\pi)\log |\log\rho {a_2^2}| \Bigr. \nonumber \\ 
&&\qquad\qquad\qquad\qquad\Bigl. + \log^24+\log 16 (1+\log \pi )+\log \pi (2+\log \pi ) -1 -2C \Bigr) \Bigg\rbrack \ .
\label{eq:conven1}
\end{eqnarray}
One readily finds the energy density in the second convention by choosing $a_2=a e^\gamma/2$ and $\lambda = e^{-2\gamma}/\pi^2$
in eq.~\ref{eq:schick}.

\section{Conclusion}

\noindent In this paper we have computed the ground-state energy of
${\bf N}$ identical bosons which interact via the most general
finite-range potential in a finite area. This energy is expressed as a
double perturbative expansion in the two-body interaction strength,
which is logarithmically dependent on the system size $L$, as well as
in inverse powers of $L$ by way of operators that break scale
invariance at the classical level. Effective range corrections and the
leading effect of three-body forces enter at ${\cal O}(L^{-4})$. The
structure of the expansion is largely dictated by scale invariance and
its logarithmic breaking. Indeed, the EFT with the leading two-body
interaction acts very much like a renormalizable field theory with a
coupling constant that runs logarithmically. All other interactions
beyond the leading two-body interaction generate power-law breaking of
scale invariance. Using the finite-area ground-state energy as a
starting point, we have also explicitly evaluated the sums that diverge
with powers of ${\bf N}$ and recovered the well-known low density
expansion of the ground-state energy density in the thermodynamic
limit.

We have seen in this paper that the many-body boson system in two
spatial dimensions is significantly simpler from a mathematical
standpoint than its counterpart in three spatial dimensions. The
tractability of the two-dimensional system is due both to the
logarithmically broken scale invariance of the system at leading order
in the momentum expansion in the two-body sector, as well as due to
the expression of two-dimensional lattice sums as products of familiar
one-dimensional sums. These two features allow one to move smoothly
between two weakly-coupled quantum regimes that are related by
infinite resummations. In particular, this tractability allows one to
calculate the leading and sub-leading finite-size corrections to the
thermodynamic limit equation-of-state. In principle, this will enable
the quantification of finite-size effects in experimental results
involving ultra-cold atoms interacting in two spatial dimensions. With
the results found in this paper, it would be interesting to
investigate the transition between the confined and
thermodynamic-limit regimes using quantum Monte-Carlo methods.

It should be clear that the method presented here for computing the
equation-of-state and low-density properties of the Bose gas in the
thermodynamic limit is not particularly efficient. Indeed, the
technology developed in Ref.~\cite{braa} and carried out in the
two-dimensional case in Ref.~\cite{Andersen} provides the most
efficient and sensible method for treating the low-density limit in a
model-independent way. Nevertheless, it is interesting to see that the
results of the low-density quantum loop expansion can be obtained in
an explicit model-independent construction without any reference to
mean field theory.

\section*{Acknowledgments}

\noindent I wish to thank Will Detmold, David Kaplan and Martin Savage
for useful discussions and for allowing me access to unpublished
notes. This work was supported in part by NSF CAREER Grant
No. PHY-0645570.


\renewcommand{\theequation}{A-\arabic{equation}}
\setcounter{equation}{0}  

\section*{\hspace{-1.1cm} APPENDIX I: Two-dimensional lattice sums}  

\noindent In an even number of dimensions, it is possible to decompose multidimensional lattice sums
into products of simple sums~\cite{Hardy,Glasser,Zucker} using methods pioneered by Jacobi~\cite{Jacobi}.
For instance, of particular interest to this paper is the sum
\begin{eqnarray}
{\cal P}_{2s}\ \equiv\ \sum^\infty_{{\bf n}\in\mathbb{Z}^2\neq 0} \frac{1}{ {(\bf n}^2)^s} \ = \ 4 {\bf \zeta}(s){\bf \beta}(s) \ ,
\label{eq:28}
\end{eqnarray}
which is valid for $s>1$, where
\begin{eqnarray}
{\bf \zeta}(s)\ \equiv \ \sum_{m=0}^\infty \frac{1}{(m+1)^s} \ , \qquad  {\bf \beta}(s)\ \equiv \ \sum_{m=0}^\infty \frac{(-1)^m}{(2m+1)^s} \ 
\label{eq:29}
\end{eqnarray}
are the Riemann zeta function and Dirichlet beta function,
respectively.  The case $s=1$ that we are interested in is singular as
$\zeta(s)$ has a simple pole at $s=1$.  In order to subtract this pole,
we require the Laurent expansions of ${\bf \zeta}(s)$ and ${\bf
\beta}(s)$ about $s=1$~\cite{Weisstein}.  We have
\begin{eqnarray}
{\bf \zeta}(s)\ &=&\ \frac{1}{s-1}\ +\ \gamma \ +\ {\mathcal O}\left( s-1 \right) \ ; \nonumber \\ 
{\bf \beta}(s)\ &=&\ \frac{\pi}{4}\Big\lbrack1+ (s-1)\left( \gamma + 2 \log 2 +3 \log\pi -4 \log \Gamma\left(\textstyle{\frac{1}{4}}\right) \right)+\ {\mathcal O}\left( (s-1)^2 \right)\Big\rbrack \ ,
\label{eq:30}
\end{eqnarray}
where $\gamma$ is Euler's constant. One then easily finds 
\begin{eqnarray}
{\cal P}_2\ \equiv\ \lim_{s\rightarrow 1}\left( \sum^{\infty}_{{\bf n}\neq 0} \frac{1}{ ({\bf n}^2)^s} \ - \ \frac{\pi}{s-1}\right)\ = \ 4\pi \log\left(e^{\frac{\gamma}{2}} \pi^{-\frac{1}{4}}\Gamma\left(\textstyle{\frac{3}{4}}\right)\right) \ .
\label{eq:35}
\end{eqnarray}

It is straightforward to generalize eq.\ref{eq:28} to 
\begin{eqnarray}
\sum^\infty_{{\bf n}\neq 0} \frac{1}{ {(\bf n}^2 - \eta^2)^s}\ =\ 
4\sum_{\ell=0}^\infty \frac{(-1)^\ell}{(2\ell +1)^s}\;\zeta\left(s, 1 - \frac{\eta^2}{(2\ell +1)}\right) \ ,
\label{eq:28mod}
\end{eqnarray}
which is valid for $s>1$. Here $\zeta(s,a)$ is the Hurwitz zeta function~\cite{Weisstein}. 
Using methods similar to those outlined above, one readily finds
\begin{eqnarray}
\hspace{-0.4cm}
\lim_{s\rightarrow 1}\left( \sum^\infty_{{\bf n}\neq 0} \frac{1}{ {(\bf n}^2 - \eta^2)^s} \ - \ \frac{\pi}{s-1}\right)
&=& {\cal P}_2 -\pi \gamma - 4\sum_{\ell=0}^\infty \frac{(-1)^\ell}{(2\ell +1)}\;\psi_0\left( 1 - \frac{\eta^2}{(2\ell +1)}\right),
 \label{eq:40}
\end{eqnarray}
where $\psi_0$ is the digamma function. Finally, one can write
\begin{eqnarray}
{\cal S}_2\left( \eta \right)\ =\ -\frac{1}{\eta^2}\ + {\cal P}_2 \ -\pi \gamma - 4\sum_{\ell=0}^\infty \frac{(-1)^\ell}{(2\ell +1)}\;\psi_0\left( 1 - \frac{\eta^2}{(2\ell +1)}\right) \ .
\label{eq:25b}
\end{eqnarray}

\section*{\hspace{-1.1cm} APPENDIX II: Catalan sums}  

\noindent In this appendix we evaluate the sums which diverge with powers of ${\bf N}$
in the thermodynamic limit. The first sum we wish to evaluate, eq.~\ref{eq:Gdefined2}, may be expressed as
\begin{eqnarray}
{\cal G}\left(z \right)&=&\sum_{n=2}^{\infty}(-1)^n C(n)z^n{\cal P}_{2n} \ ,
\label{eq:app1a}
\end{eqnarray}
where the $C(n)$ are the Catalan numbers\footnote{$C(n-2)$ is the number of ways in which a regular 
$n$-gon be divided into $n-2$ triangles if different orientations are 
counted separately~\cite{Weisstein} (Euler's polygon division problem). They are related to the central
binomial coefficients via $C(n)={2n \choose n}/(n+1)$.}, which have the integral representation~\cite{sofo}
\begin{eqnarray}
C(n)\ =\ \frac{4}{\pi} \int_0^\infty \frac{\omega^2}{(1+\omega^2)^2}\left( \frac{4}{(1+\omega^2)}\right)^n d\omega\ .
\label{eq:app1b}
\end{eqnarray}
Using eq.~\ref{eq:28} we can write
\begin{eqnarray}
{\cal G}\left(z \right)&=&   
\frac{16}{\pi} \int_0^\infty \frac{d\omega\,\omega^2}{(1+\omega^2)^2}\sum_{n=2}^{\infty}{\bar z}^n \xi(n)\beta(n) \ ,
\label{eq:app1c}
\end{eqnarray}
where ${\bar z}\equiv -4z/(1+\omega^2)$. 
By expanding eq.~\ref{eq:40} and comparing with eq.~\ref{eq:28}, it is straightforward to find
\begin{eqnarray}
\sum_{n=2}^{\infty}{\bar z}^n \xi(n)\beta(n)\ =\ {\bar z}\left( -\frac{\pi\gamma}{4}\ -\ 
\sum_{\ell=0}^\infty \frac{(-1)^\ell}{(2\ell +1)}\;\psi_0\left( 1 - \frac{\bar z}{(2\ell +1)}\right) \right).
\label{eq:app1d}
\end{eqnarray}
Using the asymptotic form of the digamma function for large argument as well as the Dirichlet sums~\cite{Weisstein}
\begin{eqnarray}
\beta(0)\ =\ \frac{1}{2}\quad ;\quad \beta'(1)\ =\ -\sum_{\ell=0}^{\infty} \frac{(-1)^\ell}{\left(2\ell+1\right)} \log\left(2\ell+1\right) \ =\ -\frac{1}{4}\left(\pi \gamma \ -\ {\cal P}_2 \right) \ ,
\label{eq:app1e}
\end{eqnarray}
one finds
\begin{eqnarray}
{\cal G}\left(z \right)&=&   \pi z \left( \log z + \frac{1}{2} + \frac{{\cal P}_2}{\pi} \right) \ +\ 1  \ +\ {\cal O}(z^{-1}) \ .
\label{eq:app1f}
\end{eqnarray}
And finally, matching to eq.~\ref{eq:Gdefined3FS},
\begin{eqnarray}
{\mathfrak g}(z) &=&  \pi \log z\ +\ \frac{\pi}{2}\ +\ {\cal P}_2 \qquad ; \qquad {\mathfrak g}_0  \ =\ 1 \ .
\label{eq:gdefined4APP} 
\end{eqnarray}

Similarly, eq.~\ref{eq:Idefined2} may be written as
\begin{eqnarray}
{\cal H}\left(z \right)&=& \sum_{n=2}^{\infty}(-1)^{n+1}(n+1)C(n)z^{n+1}{\cal P}_{2n} \ .
\label{eq:app1g}
\end{eqnarray}
Using the integral representation~\cite{Weisstein}
\begin{eqnarray}
(n+1)C(n)\ =\ \frac{2}{\pi} \int_0^\infty \frac{1}{(1+\omega^2)}\left( \frac{4}{(1+\omega^2)}\right)^n d\omega\ ,
\label{eq:app1h}
\end{eqnarray}
and proceeding as above one finds
\begin{eqnarray}
{\cal H}\left(z \right)&=&   -2 \pi z^2 \left( \log z + 1 + \frac{{\cal P}_2}{\pi} \right) \ -\ z \ +\ {\cal O}(z^0) \ .
\label{eq:app1i}
\end{eqnarray}
And finally,
\begin{eqnarray}
{\mathfrak h}(z) &=&   -2\pi \log z\ -\ 2\pi\ -\ 2{\cal P}_2 \qquad ; \qquad {\mathfrak h}_0  \ =\ -1 \ .
\label{eq:hdefined4APP}
\end{eqnarray}

As the convergence properties of the Catalan sums determine the transition to the low density regime, we can
get a more accurate measure of the expected region of validity of eq.~\ref{eq:Nboson3} through a simple
convergence test. According to D'Alembert's ratio test, the sum, eq.~\ref{eq:app1a}, is convergent
when
\begin{eqnarray}
\lim_{n\rightarrow\infty}\frac{C(n+1){\cal P}_{2n+2}}{C(n){\cal P}_{2n}}z\ < \ 1 \ ,
\label{eq:DAlem}
\end{eqnarray}
from which one easily finds $z<1/4$, or ${\bf N}\alpha_2' <\pi^2/4$. The same estimate follows from the
other sum, eq.~\ref{eq:app1g}.

\vfill\eject


\begin{thebibliography}{100}

\bibitem{Hadz}
Z.~Hadzibabic {\sl et al.},
Nature {\bf 441}, 1118 (2006).

\bibitem{clade}
P.~Clad\'e {\sl et al.}, 
arXiv:0805.3519 (2008).

\bibitem{posa}
A.~Posazhennikova, Rev. Mod. Phys. {\bf 78}, 1111 (2006).

\bibitem{Bloch:2008zz}
I.~Bloch, J.~Dalibard and W.~Zwerger,
Rev.\ Mod.\ Phys.\  {\bf 80}, 885 (2008).

\bibitem{Mermin:1966fe}
  N.~D.~Mermin and H.~Wagner,
  Phys.\ Rev.\ Lett.\  {\bf 17}, 1133 (1966).

\bibitem{Hohenberg:1967zz}
  P.~C.~Hohenberg,
  Phys.\ Rev.\  {\bf 158}, 383 (1967).

\bibitem{Kosterlitz:1973xp}
  J.~M.~Kosterlitz and D.~J.~Thouless,
  J.\ Phys.\ C C {\bf 6}, 1181 (1973).

\bibitem{fishe}
D.S.~Fisher and P.C.~Hohenberg,
Phys. Rev. B {\bf 37}, 4936 (1988).

\bibitem{Maiani:ca}
L.~Maiani and M.~Testa,
{\it Phys. Lett.} {\bf B245}, 585 (1990).

\bibitem{Luscher:1986pf}
  M.~L\"uscher,
  Commun.\ Math.\ Phys.\  {\bf 105}, 153 (1986).

\bibitem{Luscher:1990ux}
M.~L{\"u}scher,
{\it Nucl. Phys.} {\bf B354}, 531 (1991).

\bibitem{Beane:2003da}
  S.~R.~Beane, P.~F.~Bedaque, A.~Parre\~no and M.~J.~Savage,
  Phys.\ Lett.\  B {\bf 585}, 106 (2004).

\bibitem{Beane:2007qr}
  S.~R.~Beane, W.~Detmold and M.~J.~Savage,
  Phys.\ Rev.\  D {\bf 76}, 074507 (2007).

\bibitem{Tan:2007bg}
  S.~Tan,
  Phys.\ Rev.\  A {\bf 78}, 013636 (2008).

\bibitem{Detmold:2008gh}
  W.~Detmold and M.~J.~Savage,
  Phys.\ Rev.\  D {\bf 77}, 057502 (2008).

\bibitem{Beane:2007es}
  S.~R.~Beane {\sl et al.}, 
  Phys.\ Rev.\ Lett.\  {\bf 100}, 082004 (2008).

\bibitem{Fiebig:1994qi}
  H.~R.~Fiebig, A.~Dominguez and R.~M.~Woloshyn,
  Nucl.\ Phys.\  B {\bf 418}, 649 (1994).

\bibitem{schic}
M.~Schick, Phys. Rev. A {\bf 3}, 1067 (1971).

\bibitem{popo} 
V.N.~Popov, Theor. Math. Phys. {\bf 11}, 565 (1972).

\bibitem{chern} 
A.Yu.~Cherny and A.A.~Shanenko, Phys. Rev. E {\bf 64}, 027105 (2001).

\bibitem{Andersen}
J.O.~Andersen
  Eur.\ Phys.\  J. {\bf B28}, 389 (2002).

\bibitem{Braaten:2000eh}
  E.~Braaten, H.~W.~Hammer and S.~Hermans,
  Phys.\ Rev.\  A {\bf 63}, 063609 (2001).

\bibitem{Kaplan:1998we}
  D.~B.~Kaplan, M.~J.~Savage and M.~B.~Wise,
  Nucl.\ Phys.\  B {\bf 534}, 329 (1998).

\bibitem{vanKolck:1998bw}
U.~van Kolck,
{\it Nucl. Phys.} {\bf A645}, 273 (1999).

\bibitem{Kaplan:2005es}
  D.~B.~Kaplan,
  arXiv:nucl-th/0510023; {\it Unpublished Notes}.

\bibitem{Jackiw:1991je}
  R.~Jackiw,
  in *Jackiw, R.: Diverse topics in theoretical and mathematical physics*, 35-53. 

\bibitem{Hammer:2004as}
  H.~W.~Hammer and D.~T.~Son,
  Phys.\ Rev.\ Lett.\  {\bf 93}, 250408 (2004).

\bibitem{Luscher:1990ck}
  M.~Luscher and U.~Wolff,
  Nucl.\ Phys.\  B {\bf 339}, 222 (1990).

\bibitem{Detmold:UN}
  W.~Detmold and M.~J.~Savage,
{\it Unpublished Notes}.

\bibitem{Lee:1957zzb}
  T.~D.~Lee, K.~Huang and C.~N.~Yang,
  Phys.\ Rev.\  {\bf 106}, 1135 (1957).

\bibitem{braa}
E.~Braaten and A.~Nieto, Eur.\ J.\ Phys. B.\ {\bf 11}, 143 (1999)

\bibitem{astrNU}
G.E.~Astrakharchik {\sl et al.}, 
Phys. Rev. A {\bf 81}, 013612 (2010).

\bibitem{Brodsky:1982gc}
  S.~J.~Brodsky, G.~P.~Lepage and P.~B.~Mackenzie,
  Phys.\ Rev.\  D {\bf 28}, 228 (1983).

\bibitem{Brodsky:1994eh}
  S.~J.~Brodsky and H.~J.~Lu,
  Phys.\ Rev.\  D {\bf 51}, 3652 (1995).

\bibitem{astrDC}
G.E.~Astrakharchik {\sl et al.}, 
Phys. Rev. A {\bf 79}, 051602(R) (2009).

\bibitem{moraDC}
C.~Mora and Y.~Castin, 
Phys. Rev. Lett. {\bf 102}, 180404 (2009).

\bibitem{Glasser}
 M.L.~Glasser, J.\ Math.\ Phys.\ {\bf 14}, 409 (1973).
 
\bibitem{Zucker}
I.J.~Zucker, J.\ Math. \ Phys. A. \ {\bf 7}, 1568 (1974).

\bibitem{Hardy}
G.H.~Hardy, Mess.\ Math.\ {\bf 49}, 85 (1919).

\bibitem{Jacobi}
C.G.I.~Jacobi, {\it Fundamenta Nova Theoriae Functionum Ellipticarum}, (Konigsberg).

\bibitem{Weisstein}
E.W.~Weisstein, http://mathworld.wolfram.com/

\bibitem{sofo}
A.~Sofo, JIPAM {\bf 10}, 69 (2009).


\end{thebibliography}
\end{document}